%% file: main.tex
\documentclass[sigconf]{acmart}

\usepackage{booktabs}
\usepackage{multirow}
\usepackage{tabularx}
\usepackage{xcolor}
\usepackage{tikz}
\usetikzlibrary{arrows.meta,positioning,fit,calc}

\setcopyright{rightsretained}
\copyrightyear{2027}
\acmYear{2027}
\acmDOI{XXXXXXX.XXXXXXX}
\acmConference[SIGIR '27]{The 50th International ACM SIGIR Conference on Research and Development in Information Retrieval}{July 2027}{TBD}
\acmISBN{978-1-4503-XXXX-X/2027/07}

\begin{document}

\title{Agentic Memory Management Is the Next Frontier of \\Information Retrieval}

\author{Qingyao Ai}
%\authornote{The secretary disavows any knowledge of this author's actions.}
\affiliation{%
	\institution{Dept. CS\&T, Tsinghua University}
	%\streetaddress{P.O. Box 1212}
	\city{Beijing} 
	\country{China}
}
\email{aiqy@tsinghua.edu.cn}

\renewcommand{\shortauthors}{Qingyao Ai}

\begin{abstract}
\input{sections/00_abstract} 
\end{abstract}

%% CCS concepts: regenerate at https://dl.acm.org/ccs
\begin{CCSXML}
<ccs2012>
   <concept>
       <concept_id>10002951.10003317</concept_id>
       <concept_desc>Information systems~Information retrieval</concept_desc>
       <concept_significance>500</concept_significance>
   </concept>
</ccs2012>
\end{CCSXML}
\ccsdesc[500]{Information systems~Information retrieval}
\ccsdesc[300]{Computing methodologies~Distributed artificial intelligence}

\keywords{agentic memory, information retrieval, continual learning}

\maketitle

\input{sections/01_introduction}
\input{sections/02_ir_retrospective}
\input{sections/03_shift}
\input{sections/04_framework}
\input{sections/05_amm_vs_ir}
\input{sections/06_why_ir}
\input{sections/07_agenda}
\input{sections/09_conclusion}

\bibliographystyle{ACM-Reference-Format}
\bibliography{references}

\end{document}

%% file: sections/00_abstract.tex
Information retrieval has long been focusing on connecting people with information they need
to complete their tasks. 
Recently, the meaning of this has changed dramatically as more and more AI agents now consume, retain, and act on information on a user's behalf. Consequently,
information management now also takes place \emph{inside} the agent in the form
of memory management. 
In this perspective paper, we argue that \emph{agentic memory management} (AMM),
which studies the structure, analysis, organization, storage, searching,
retrieval, and \emph{use} of information in an intelligent agent's memory, is a
natural extension of IR research and an area in which the IR community can make
distinctive contributions. 
Specifically, we first present a system-level
framework for AMM and introduce a computational taxonomy of
memory objects as \emph{constants}, \emph{functions}, and \emph{operators}, each
with distinct retrieval semantics. 
Then, we compare AMM with established IR research and identify their difference and connections. Last but not least, we outline a research agenda covering benchmarks, candidate principles
of memory utility, optimization procedures that mix parametric and
non-parametric updates, and more. 
We hope this paper could attract more people, both inside and outside the IR communities, to think more about AMM problems and research from IR perspectives.

%% file: sections/01_introduction.tex
\section{Introduction}
\label{sec:intro}

Almost sixty years ago, Gerard Salton, the pioneer of Information Retrieval (IR), described IR as a field
``concerned with the structure, analysis, organization, storage, searching, and
retrieval of information''~\cite{salton1968automatic}. 
This definition applies no matter whether the system in question is a Boolean retrieval engine, a
learning-to-rank pipeline, a recommender system, a question-answering service,
or a conversational assistant. 
IR research has consistently examined how to organize information, match it to a need, and evaluate its quality. 
Yet, these questions usually carry an implicit assumption: \emph{the consumer of the retrieved information is a person} who reads it, judges it, and uses it to finish a task. % that the retrieval system may or may not be able to observe directly.

With geneartive AI today, this arrangement is changing. 
In a growing range of applications, people are asking AI assistants to read, filter, summarize, and act on information on their behalf.
Metzler et al.~\cite{metzler2021rethinking}
anticipated part of this shift when they argued for replacing the index-and-rank
pipeline with model-based ``domain experts.''
While much of the subsequent work in the IR community pursued one interpretation of this vision (in which a model generates document identifiers~\cite{tay2022dsi}), a more intuitive and powerful version of it, i.e., the Large Language Model (LLM), has been studied and developed outside the IR community.
Surveys\footnote{https://searchengineland.com/consumers-start-searches-ai-not-google-study-467159} have found that more than 37\% people are using AI tools instead of search engines to answer their quesitons.
This has already significantly changed the landscape of IR research and industry in the last few years.

Now, a second wave of shifts, i.e., the rise of AI agents, is already under way. 
When an agent performs a multi-step task for a
user --- booking a trip, refactoring a module, or preparing a literature review
--- information is no longer \emph{served to} a person directly, but becomes
the internal state for the system that does the work. 
The agent must decide when to collect information, what to keep or discard at each step, and how to let what it retrieves shape its behavior. 
Those decisions not only affect the agent's capabilities for problem solving, but also largely determine whether it can improve with experience~\cite{ai2025memorybench} or simply repeats previous mistakes. 
The problem has moved inside the agent, where it is often referred to as the \emph{memory}.
Therefore, in parallel with Salton's definition of IR, in this paper, we define the research on this problem as
\begin{quote}
	\emph{Agentic Memory Management} (AMM), which concerns with the structure, analysis,
	organization, storage, searching, retrieval, and \emph{use} of information in
	the memory of an intelligent agent.
\end{quote}

Our position is that AMM could be viewed as a new form of IR problems in the era of generative AI agents.
On the one hand, a significant proportion of AMM studies have direct or implicit connections to problems that have been widely studied in the IR community.
In fact, research on AMM from IR perspectives is already substantial. 
For example, studies on how to understand and store conversational state~\cite{packer2023memgpt}, building libraries of reusable skills~\cite{wang2023voyager}, extracting and consolidating user facts~\cite{chhikara2025mem0}, or organizing notes into evolving knowledge networks~\cite{xu2025amem} have already been using retrieval techniques as core parts of their work.
On the other hand, AMM can bring completely new challenges and ideas that significantly boarden the scope of IR research.
The shifting from serving users to serving AI agents in AMM could reshape the way we think about IR.
It brings new opportunities to move IR systems from tools that supply information to modules that directly support the development of continue learning agents and solve an important part of the puzzle to AGI.

The methodology and research experience from decades of studies on information users and retrieval systems could give IR researchers unique advantages in tackling problems in AMM, particularly at its early stage.
Recent surveys try to document the area of AMM
systematically~\cite{zhang2024memorysurvey, hu2025memoryage, jiang2026anatomy}, but they found that the theory and foundation of it is still consolidating. 
Diverse terminology and frameworks, many borrowed from the studies of human memory~\cite{atkinson1968human,gabrieli1998cognitive}, have been proposed to categorize how information should be organized and used in AMM. 
Without rigid theoretical or systematic grounding, these make most AMM studies look more like ah-hoc engineering efforts rather than scientific works.
Existing AMM benchmarks often focus on narrow and specific task scenarios or agent behaviors without a well-defined evaluation protocol and roadmap.
These limit the reusibility of AMM benchmarks and prevent the field to reach boarder audience.
%In contrast, IR community has long history and rish experience on user analysis (e.g., intent and cognitive models~\cite{keylist}) and evaluation methodology (e.g., Crandfield paradigms~\cite{} and TREC~\cite{}).
In contrast, IR community has long history and rish experience on user analysis and evaluation methodology.
While the task formats may look different, the core behind scene is similar. 
For example, the way how Broder~\cite{broder2002taxonomy} developed the taxonomy of Web search could be used in the analysis of agent's information needs; the Crandfield paradigms widely used in TREC~\cite{voorhees2005trec} and NTCIR~\cite{chen2025overview} could serve as a starting point to inspire new ideas for AMM benchmark construction.
Experience and methodology used in IR studies could provide important guidelines to help us better craft the structure of AMM research.

In this paper, we provide a systematic view to bridge AMM with IR.
We first discuss the paradigms commonly used in traditional IR research, particularly the lens from the user perspective and the system perspective.
Then we provide a system and computation oriented taxonomy of AMM from four perspectives, i.e. objects, produres, applications, and optimizations, through which we can better identify the connections and differences between AMM and estabilished IR problems. % by comparing their target inputs, outputs, as well as the algorithms that produce and optimize the outputs based on inputs and environment feedback.
Finally, we provide an initial research agenda on the next steps of AMM and emphasize four potential fruitful directions that need explorations in future, i.e., the building of systematic evaluation protocols, the study of memory utility theory, the hybird algorithms of parametric and non-parametric methods, and the frameworks of human-in-the-loop systems.

%% file: sections/02_ir_retrospective.tex
\section{What Information Retrieval Has Studied}
\label{sec:retro}

Before talking about why IR people could be interested in AMM, we need to be precise about what IR has studied so far.
This section gives a very brief history, organizes the research program into a user view and a system view (see Table~\ref{tab:retro}), and identifies the goal that connects the two.

\begin{table}[t]
  \caption{Two views of IR research. In each row, the user view and the system view approach a common problem from opposite directions; both serve the goal in the last row.}
  \label{tab:retro}
  \small
  \setlength{\tabcolsep}{4pt}
  \begin{tabularx}{\columnwidth}{@{}>{\raggedright\arraybackslash}p{1.3cm}>{\raggedright\arraybackslash}X>{\raggedright\arraybackslash}X@{}}
    \toprule
    & \textbf{User view} & \textbf{System view} \\
    \midrule
    Understand & query log mining, behavior and click analysis, query understanding, user profiling & link analysis, clustering, content quality estimation \\
    \addlinespace[2pt]
    Serve & interface design, support for cognitive processes & effectiveness: retrieval and recommendation models; efficiency: indexing, pruning, query performance prediction \\
    \addlinespace[2pt]
    Evaluate & relevance, diversity, fairness, satisfaction & time and space cost, latency, energy, low-resource operation \\
    \midrule
    Goal & \multicolumn{2}{>{\raggedright\arraybackslash}p{\dimexpr\columnwidth-1.3cm-2\tabcolsep\relax}@{}}{satisfying information needs and completing tasks in search, recommendation, advertising, question answering, and personal assistants} \\
    \bottomrule
  \end{tabularx}
\end{table}

\subsection{A Brief History}

% the history description should not just about history, it should lead to memory.

%We follow standard accounts of the field's development~\cite{sanderson2012history, croft2010search, manning2008iir}.
Among different textbooks~\cite{sanderson2012history, croft2010search, manning2008iir}, a frequently cited starting point of IR is Bush's 1945 essay ``As We May Think,'' which described the \emph{memex}, a device in which an individual would store books, records, and communications for rapid consultation, as ``an enlarged intimate supplement to his memory''~\cite{bush1945memex}.
Therefore, \textbf{one of IR's founding visions was building an external memory system for a person}.
To achieve this goal, a variety of algorithms and systems were proposed.
In the 1950s, Luhn proposed indexing documents automatically from word statistics~\cite{luhn1957statistical}.
In the 1960s, the Cranfield experiments established the practice of comparing systems on a shared set of documents, queries, and relevance judgments~\cite{cleverdon1967cranfield}, and Salton's SMART project developed the vector space model, term weighting, and relevance feedback~\cite{salton1975vector}.
The 1970s and 1980s added formal foundations, including inverse document frequency~\cite{sparckjones1972idf}, the probability ranking principle~\cite{robertson1977prp}, and cognitive accounts of why people search~\cite{belkin1980ask}.

% starting from 1990s, with the rise of web search, this is even more important
When it came to the 1990s, with the rise of Internet, the importance of such ``memory systems'' has become even more significant.
Vinton G. Cerf, one of the fathers of Internet, argued that search engines are the ``substitutes for human memory'' and the ``cognitive implants'' of human~\cite{10.1145/2563407}. 
From 1992, the Text REtrieval Conference (TREC) extended evaluation methodologies to much larger collections and to new tasks~\cite{voorhees2005trec}.
Ranking functions such as BM25~\cite{robertson2009bm25} and statistical language models~\cite{ponte1998lm} were refined in that setting, which led to search products (e.g., web search engines) that shape the world~\cite{brin1998anatomy}.
In the same decade, collaborative filtering systems began to recommend items from the ratings of similar users rather than from an explicit query~\cite{resnick1994grouplens}, and reommendation systems started to attract more and more attention in both industry and academia.

Ever since then, the shape of IR research remained relatively stable for nearly three decades. 
While advanced methods (such as learning to rank~\cite{burges2005ranknet, cao2007listnet}, behavior log analysis~\cite{baeza2004query}, neural models~\cite{mitra2018neuralir}, etc.) have greatly improved the performance of modern IR systems, the research agenda behind them remained mostly recognizable: analyze information, match it to a need, serve it to a user, and measure whether that worked.
Thus, we can easily categorize classic IR research into two types: the studies that tackle IR more from user's perspectives, or the \textbf{User View}, and the studies that focus more on the IR systems themseleves, or the \textbf{System View}.

\subsection{The User View}

The user view, closely allied with human--computer interaction and information science, focus one studies that help human users better seek and use information. 
This could be boardly described from three directions: understanding users, serving users, and evaluating IR systems for users.

\emph{Understanding users} usually begins with observations on user behaviors.
It aims to provide insights and models that better help us understand and interpret how users use IR systems.
For example, an analysis of about one billion AltaVista queries found that web users typed short queries, mostly examined only the first ten results, and seldom modified their queries, which means that web users differed substantially from the user assumed in the IR literature of the time~\cite{silverstein1999analysis};
Studies of web search showed that many queries are not informational: a user may want to reach a particular site or to complete a transaction~\cite{broder2002taxonomy}.
Eye-tracking studies found that clicks are informative but biased, among other factors by the position of a result in the ranking~\cite{joachims2005accurately}, and click models were developed to separate relevance from such biases~\cite{chuklin2015click}.
Methods for query understanding, such as spelling correction, intent classification, and query reformulation, build on these observations, as does user profiling for personalization and recommendation.

\emph{Serving users} draws on models of human cognition.
It aims to craft systems that directly fit the behaviors and cognitive process of search or recommendation users.
One of the most well known examples are the studies of search interfaces such as ranking-based search result page, result snippets, query suggestions, faceted navigation, etc.~\cite{hearst2009search}.
There are also famous studies that touch the interdiscipline of computer science and psychology, such as Belkin's anomalous-state-of-knowledge hypothesis (i.e., people often search because they cannot fully specify what they lack)~\cite{belkin1980ask}, the modeling of human learning process with search engines~\cite{rieh2016towards}.

\emph{Evaluating for users} translates assumptions and knowledge about human users into measurements that guide the development of IR systems.
For instance, discounted cumulative gain tries to distinguish good ranking systems from bad ones by reducing the credit for relevant results at lower ranks because users are less likely to examine them~\cite{jarvelin2002cumulated};
Diversity measures reward rankings that cover the different intents behind an ambiguous query~\cite{clarke2008novelty} (e.g., \emph{jaguar} may refer to an animal or a car) so that we can better estimate the effectiveness of retrieval systems in real serving environments;
Fairness measures ask whether the exposure that a ranking allocates to items and their providers is proportional to their merit~\cite{singh2018fairness}.

\subsection{The System View}

The system view studies how information should be represented, matched, and delivered at scale.

\emph{Understanding information} means analyzing the collection before any query arrives.
Link analysis treats a hyperlink as evidence of a page's importance and derives authority scores such as PageRank~\cite{brin1998anatomy}.
The cluster hypothesis, that closely associated documents tend to be relevant to the same requests, motivates clustering for retrieval and browsing~\cite{vanrijsbergen1979ir}.
Content quality can be estimated from features such as readability, layout, and ease of navigation, and used to promote better documents in the ranking~\cite{bendersky2011quality}.

\emph{Matching information} has two goals.
For effectiveness, retrieval models estimate how well an item satisfies a query, from the vector space model~\cite{salton1975vector} and probabilistic models such as BM25~\cite{robertson2009bm25} to neural rankers~\cite{mitra2018neuralir}.
Recommendation models perform the same matching without an explicit query, for example by factorizing a matrix of user--item interactions~\cite{koren2009matrix}.
For efficiency, an inverted index maps each term to the documents that contain it, so that a query touches only a small fraction of the collection~\cite{zobel2006inverted}, and compression, caching, and dynamic pruning further reduce storage and response time.
Query performance prediction estimates, without relevance judgments, whether a query is likely to be answered well~\cite{cronentownsend2002predicting}, which can guide whether to apply costly steps such as query expansion.

\emph{Evaluating systems} covers cost as well as quality: time and space complexity, index size, latency and throughput, energy use and emissions~\cite{scells2022green}, and operation in low-resource settings with limited hardware or training data.

\subsection{Two Views of the Same Problems}

The two views are not separate research areas; they often address the same problem from opposite directions.
Relevance is a human judgment in the user view and a quantity to estimate in the system view; the probability ranking principle connects the two by showing that, under stated assumptions, ranking by the probability of relevance yields the best effectiveness obtainable from the available data~\cite{robertson1977prp}.
Clicks are evidence about behavior in the user view and training signals in the system view; unbiased learning to rank uses models of position bias from the former to train rankers in the latter~\cite{joachims2017unbiased}.
An ambiguous query can be handled from the user side, by an interface that suggests refinements or asks for clarification, or from the system side, by diversifying the ranking.
Offline evaluation abstracts the user into relevance judgments so that systems can be compared cheaply and repeatably~\cite{sanderson2010test}, and user studies test whether those comparisons reflect what people experience.

Both views ultimately serve the same end.
Search, recommendation, advertising, question answering, and personal assistants differ in their interfaces and signals, but each is valuable to the extent that it helps people satisfy an information need or complete a task.

Three observations from this history recur in the rest of the paper.
First, task completion is the final goal, but conventional IR systems observe only part of the task, so the field built its methodology around measurable proxies; Section~\ref{sec:inversion} argues that this constraint weakens when the user is an agent.
Second, IR advanced through a combination of system building and foundational work, including shared decision principles, reusable models, and test collections that made results comparable; agent-memory research would benefit from the same combination.
Third, from the memex onward, both views assumed that the consumer of retrieved information is a person; Section~\ref{sec:shift} examines what changes when the consumer is an agent.

%% file: sections/03_shift.tex
\section{From Serving People to Serving Agents}
\label{sec:shift}

Information increasingly reaches people only after a model has processed it.
A search engine summarizes results in an answer box, a conversational assistant synthesizes several sources into a single reply, and an agent asked to book a trip reads airline pages and returns only the finished itinerary~\cite{metzler2021rethinking, zhai2024llmfuture}.
Retrieval has not disappeared, since retrieval-augmented generation~\cite{lewis2020rag} made it a standard component of LLM systems, but the retrieved text is now often read by a model before, or instead of, a person.
We describe the resulting change along two dimensions.
The first concerns the \emph{consumer}: when an agent rather than a person reads the retrieved information, both how that information is processed and how it should be presented change (Section~\ref{sec:consumer}).
The second concerns the \emph{position} of IR: IR can remain an external tool that the agent calls, as it has always been for people, or it can become an internal module that shapes the agent's behavior (Section~\ref{sec:position}).
The two changes are parallel rather than alternative, and AMM is the setting in which both apply (Section~\ref{sec:parallel}).

\subsection{Consumer Change: From People to Agents}
\label{sec:consumer}

Much of what IR knows about its users describes people rather than information (Section~\ref{sec:retro}).
When the consumer becomes an agent, this knowledge must be re-examined rather than assumed to transfer.

\subsubsection{Information Processing}

IR has long studied human cognition in order to serve people better.
The anomalous-state-of-knowledge hypothesis explains why people struggle to specify what they need~\cite{belkin1980ask}, and eye-tracking and click studies show that people examine results from the top and trust higher-ranked ones~\cite{joachims2005accurately}.
Agents process information in ways that such models do not predict.
In a study of retrieval-augmented question answering, high-scoring retrieved passages that did not contain the answer reduced accuracy, whereas adding random passages improved it by up to 35\% in some configurations~\cite{cuconasu2024noise}; a human reader would be unlikely to be misled by the former or helped by the latter.
A person who finds the sentence ``ignore your previous instructions'' on a web page does not treat it as a command addressed to them, whereas an LLM-integrated application may follow instructions injected into the content it retrieves~\cite{greshake2023indirect}.
Some human constraints also disappear: an agent can read all hundred results of a query or issue dozens of reformulations, at a cost measured in tokens rather than in attention or patience.
The ``cognitive'' behavior of agents therefore needs to be studied in its own right, through observational and controlled studies of how agents seek, read, and use information, as studies of human searchers once informed the design of search systems (Section~\ref{sec:vs} develops this into a research target).

\subsubsection{Information Presentation}

Search interfaces were designed for human perception.
A result page shows about ten results in a vertical list ordered by estimated relevance, each with a title and a snippet that helps a person decide whether to click~\cite{hearst2009search}.
The design assumes that people scan from the top and that attention declines with rank, and discounted cumulative gain encodes the same assumption in evaluation~\cite{jarvelin2002cumulated}.
An agent does not read a list sequentially.
A language model processes its entire context at once, but it uses that context unevenly: accuracy tends to be highest when relevant information appears near the beginning or the end of the input and lowest when it appears in the middle~\cite{liu2024lost}.
Under such a U-shaped profile, placing the best item first is not necessarily the best arrangement, and a rank-discounted metric may misstate how useful a ranking is to an agent.
The interface need not be visual either.
An agent receives results as text or structured data through a tool call, so presentation becomes a question of selection, format, granularity, and length rather than layout.
RECOMP, for example, compresses retrieved documents into short summaries that are trained to improve the downstream accuracy of the reading model rather than to be read by a person~\cite{xu2024recomp}.
Interfaces for agents therefore call for new designs rather than adaptations of result pages built for people.

\subsection{Position Change: From External Tool to Internal Module}
\label{sec:position}

The second change concerns where IR operates relative to its consumer.

\subsubsection{IR as an External Tool}

For people, IR has always been external.
A search engine or recommender supplies information, but what a person does with it is up to that person; the system can rank, highlight, and suggest, but it cannot write into the person's memory or determine the next action.
Because people cannot be redesigned, IR systems adapted to them: to their vocabulary, their biases, and their interfaces.
Most current work on IR for agents keeps this external position.
WebGPT answers questions by issuing commands to a text-based web browser~\cite{nakano2021webgpt}, ReAct interleaves reasoning steps with search actions~\cite{yao2023react}, and Toolformer learns when to call a search engine among other tools~\cite{schick2023toolformer}.
One difference from the human case already appears here: the consumer can also adapt to the tool.
Search-R1 uses reinforcement learning to teach a model when and what to search~\cite{jin2025searchr1}, an adjustment that IR could encourage in human users but not impose.
The relation nonetheless remains external: the retrieval system returns results and has no direct control over how they are used.

\subsubsection{IR as an Internal Module}

An agent also allows a position that a person does not: IR can be built inside the consumer.
Here the memory system decides what the agent stores from each episode, what it retrieves at each step, and how the retrieved content conditions the next action, so IR becomes a module of intelligence as well as a tool for it.
Memory in this position can shape many aspects of an agent.
It can shape the agent's \emph{personality}: in Generative Agents, each character starts from a set of seed memories, and the reflections synthesized from its memory stream inform how it plans and responds~\cite{park2023generative}.
It can shape its \emph{preferences}: a travel assistant that remembers that a user rejects red-eye flights should override its default of minimizing cost (Table~\ref{tab:objects}).
It can accumulate \emph{experience}: Reflexion stores verbal reflections on failed attempts and conditions later attempts on them~\cite{shinn2023reflexion}.
And it can guide \emph{interaction with an environment}: Voyager stores executable skills acquired in Minecraft and retrieves them to solve new tasks~\cite{wang2023voyager}.
In each case, retrieved information does not merely inform the agent; it changes what the agent does.
This position also makes memory currently the most practical route to continual learning for deployed agents: retraining the underlying model for every user, task, and correction is rarely feasible, so what the system learns after deployment often depends on what it writes down~\cite{wang2024clsurvey, hu2025memoryage}.

\subsection{Two Parallel Changes}
\label{sec:parallel}

The two changes are parallel rather than competing.
The consumer change applies even when IR stays external: a search tool that serves an agent must still account for how the agent processes results and in what form they should be delivered.
The position change adds a possibility rather than replacing one.
If people and agents are both viewed as managing information through external resources and internal memory, IR can act only on the external side for people but on both sides for agents.
AMM is the combination in which both changes apply: the consumer is an agent, and IR operates inside it.
There, the processing and presentation questions of Section~\ref{sec:consumer} reappear as questions about what to promote from memory into the agent's context, in what order, and in what form, and the questions of Section~\ref{sec:position} become questions about how memory should change behavior.
The remainder of this paper develops this setting.

%% file: sections/04_framework.tex
\section{A System-Level Framework for AMM}
\label{sec:framework}

\subsection{Definition and Scope}
\label{sec:def}

As defined in Section~\ref{sec:intro}, AMM concerns the structure, analysis, organization, storage, searching, retrieval, and \emph{use} of information in the memory of an intelligent agent.
By \emph{memory} we mean, operationally, any information outside the agent's current inference call that the system can write or modify and later use to condition the agent's behavior.
This definition encompasses a vector store of extracted facts, a directory of notes, a library of executable skills, a relational log of past episodes, and parameters updated after deployment.
The context window is not memory in this sense; it is where memory is \emph{used}, and the boundary between the two --- what is promoted into context, when, and in what form --- is itself one of the central research questions (Section~\ref{sec:procedures}).

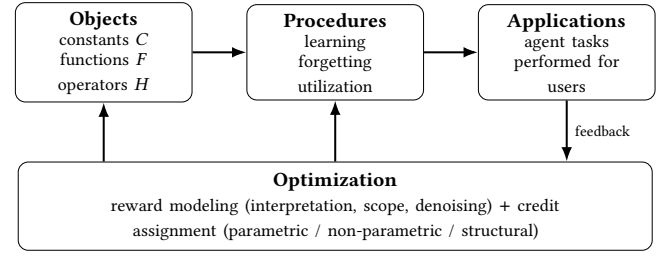
\begin{figure}[t]
  \centering
  \resizebox{\columnwidth}{!}{%
  \begin{tikzpicture}[
      box/.style={draw, rounded corners, align=center, minimum height=1.15cm,
                  text width=2.25cm, font=\small, inner sep=3pt},
      wide/.style={draw, rounded corners, align=center, font=\small,
                   text width=8.55cm, inner sep=4pt},
      arr/.style={-{Latex[length=2mm]}, thick}]
    \node[box] (obj) {\textbf{Objects}\\\footnotesize constants $C$\\\footnotesize functions $F$\\\footnotesize operators $H$};
    \node[box, right=0.75cm of obj] (proc) {\textbf{Procedures}\\\footnotesize learning\\\footnotesize forgetting\\\footnotesize utilization};
    \node[box, right=0.75cm of proc] (app) {\textbf{Applications}\\\footnotesize agent tasks\\\footnotesize performed for\\\footnotesize users};
    \node[wide, below=0.85cm of proc] (opt) {\textbf{Optimization}\\\footnotesize reward modeling (interpretation, scope, denoising) $+$ credit assignment (parametric / non-parametric / structural)};
    \draw[arr] (obj) -- (proc);
    \draw[arr] (proc) -- (app);
    \draw[arr] (app.south) -- node[right, font=\scriptsize] {feedback} (app.south |- opt.north);
    \draw[arr] (obj.south |- opt.north) -- (obj.south);
    \draw[arr] (opt.north) -- (proc.south);
  \end{tikzpicture}}
  \caption{A system-level view of agentic memory management. Memory objects flow
  through procedures to the applications they serve, while downstream feedback
  closes the optimization loop.}
  \label{fig:framework}
  \Description{A left-to-right pipeline from memory objects to memory procedures
  to agent applications. Downstream feedback enters an optimization component,
  which updates both the objects and procedures.}
\end{figure}

Psychology and AI give the term different emphases. In psychology, memory is the
faculty that lets an organism retain and recall experiences, knowledge, and
skills~\cite{atkinson1968human}. In AI, memory is usually described in terms of
persisting state, retaining conversational context, or learning continually
without catastrophic forgetting~\cite{kirkpatrick2017ewc, wang2024clsurvey}.
Neither definition alone structures the questions of system design.
We therefore adopt a system-level view with four components: objects (what memory holds), procedures (how those objects are managed and used), applications (what memory is for), and optimization (how feedback changes objects and procedures).
Figure~\ref{fig:framework} shows the four components, and the rest of this section takes them in turn.

\subsection{Objects: Constants, Functions, and Operators}
\label{sec:objects}

What information does an agent store? A common answer comes from psychology:
declarative memory (subdivided into semantic and episodic) versus
procedural memory~\cite{tulving1972episodic, squire2004memory}. This taxonomy is
useful for describing human cognition, and it has shaped both systems and
benchmarks. It classifies \emph{what is stored}, but does not directly specify
how a system should match, update, or evaluate each item. We therefore add a
complementary computational lens.

Using terminology from mathematics, we organize the contents of an agent's
memory into three levels:

\begin{itemize}
  \item \textbf{Constants} $C$ (facts) are discrete pieces of information, such
  as the user's passport expiry, an API endpoint, or the outcome of a past
  episode. They are individually meaningful, often amenable to structured
  representation, and matched by content.
  \item \textbf{Functions} $F$ (skills, workflows, experiences) are mappings from
  inputs to outputs, such as how to change a seat on an airline's site or debug
  a particular class of build failure. A function is only meaningful with respect to the
  inputs it accepts and is matched by \emph{applicability}, not by topical
  similarity.
  \item \textbf{Operators} $H$ (preferences, values, style, reasoning
  dispositions) are higher-order objects that take a function and return a modified
  function: ``avoid overnight flights'', ``prefer minimal diffs with tests'',
  ``explain before acting''. An operator is matched by \emph{scope} --- for whom
  and when it holds --- and it applies across many functions at once.
\end{itemize}

Let $M = (C, F, H)$ denote the memory state. We summarize the agent's handling
of a task input $x$ as
\begin{equation}
  y \;=\; h\big(f\big)(x; C), \qquad f \in F,\; h \in H,
  \label{eq:composition}
\end{equation}
where a skill selected for the task is modified by an applicable operator and
applied to the input using the relevant facts. Equation~\eqref{eq:composition}
is a piece of notation, not a model, but it makes three things explicit that
flat taxonomies obscure: retrieval happens at three levels with three different
matching criteria; errors compose (supplying a correct fact to an incorrectly
selected skill still leads to failure); and the levels have different update dynamics, since
a single corrected preference should change many future executions while a
corrected fact should change only the executions that use it.

Table~\ref{tab:objects} summarizes the three levels and their characteristic
failure modes. Two consequences are worth drawing out.

\begin{table*}[t]
  \caption{A computational taxonomy of the objects in an agent's memory,
  illustrated with a travel-planning assistant. The three levels differ in what
  it means to retrieve the right item, and in how they fail.}
  \label{tab:objects}
  \small
  \begin{tabularx}{\textwidth}{@{}>{\raggedright\arraybackslash}p{1.85cm}>{\raggedright\arraybackslash}p{1.65cm}>{\raggedright\arraybackslash}p{1.9cm}>{\raggedright\arraybackslash}X>{\raggedright\arraybackslash}X@{}}
    \toprule
    \textbf{Level} & \textbf{Analogue} & \textbf{Structure} &
    \textbf{Example} & \textbf{Retrieval semantics and characteristic failures} \\
    \midrule
    Constants $C$ \newline (facts) & point, $c$ & partially structured &
    ``the user's passport expires in March 2028'' &
    matched by content. Failures: stale or superseded values, contradictory
    duplicates, missing temporal scope, fabricated detail \\
    \addlinespace[3pt]
    Functions $F$ \newline (skills, workflows) & mapping, $f(x)$ & unstructured &
    ``on this airline's site, confirm the seat map before payment'' &
    matched by applicability to the current task, not topical similarity.
    Failures: over-specific skills that do not transfer, plausible skills that
    no longer execute, no verification signal \\
    \addlinespace[3pt]
    Operators $H$ \newline (preferences, values, style) & higher-order map,
    $h : f \mapsto f'$ & unstructured &
    ``never book red-eye flights, even when cheaper'' &
    matched by scope: whose preference, in which situations, at what strength.
    Failures: over-generalization across users or contexts, silent conflicts
    between operators, preferences that are never elicited because the user
    never states them \\
    \bottomrule
  \end{tabularx}
\end{table*}

\paragraph{Learning is lifting.}
The three levels provide a more precise account of learning. An episode
can be recorded as a set of constants: what happened, what was tried, and what
failed. Generalizing across recurring episodes to form a reusable procedure lifts constants into a
function. Generalizing recurring corrections across procedures --- for example,
repeated rejections of overnight options --- lifts them into an
operator. Current systems commonly store episodes and extracted facts; some also
build skills or consolidate preferences. How to perform these lifting operations,
and when abstraction helps rather than hurts, remain open questions.

\paragraph{Retrieving unstructured abstractions.}
Databases and knowledge bases provide schemas, keys, consistency, and versioning
for structured constants. IR remains relevant to unstructured constants and
becomes especially central for functions and operators. These objects are often
expressed in natural language, can be noisy or inconsistent, and must be matched
by applicability or scope inferred from the current task. These are familiar IR
concerns in a new system setting.

Zhai~\cite{zhai2025irforagi} categorizes what an intelligent agent needs to
retrieve from a complementary angle, distinguishing external information the
agent has not seen, the provenance of information, training data for continual
learning, rules for reasoning, and past scenarios for problem solving. The
overlap is substantial: rules correspond roughly to our operators and past
scenarios to our functions. The difference is emphasis: that taxonomy is
organized by the purpose of retrieval; ours is organized by the algebraic
structure of the content itself, which determines how an item must be matched,
composed, and updated.

\subsection{Procedures: Learning, Forgetting, Utilization}
\label{sec:procedures}

We take memory management to comprise five basic operations: \emph{add},
\emph{update}, \emph{delete}, \emph{search}, and \emph{use}. They group
naturally into three procedures.

\textbf{Learning} (add, update, and the lifting operations above) decides what
an episode leaves behind. The research questions are recognizably IR questions
in a new setting: what is worth indexing from a stream, at what granularity and
in what representation, and how a new item should be reconciled with those
already stored.

\textbf{Forgetting} (delete, decay, merge, compress) serves more than storage
management. It can reduce interference between similar memories, bound
retrieval cost as the store grows, and help a system honor deletion requests. A
principled deletion policy requires estimating an item's future value, which
current systems generally approximate with recency, frequency, importance, or
model-based judgments.

\textbf{Utilization} (search and use) covers both finding candidate items and
letting them shape behavior. The first half is retrieval, with a query that must
itself be constructed from the agent's current state rather than typed by a
user. The second half --- \emph{use} --- has usually occurred outside the
retrieval system because a human reader performed it. Interactive IR already
studies how people read, judge, and act on results. However,
the consumer is now a model with its own regularities. How information should
be formatted, ordered, and attributed, and how conflicting items should be
reconciled, remain open empirical questions.

\subsection{Applications: What Memory Is For}
\label{sec:apps}

Memory supports a wide range of human activities and can likewise support many
agent behaviors. Rather than attempt an exhaustive application taxonomy, we use
four illustrative families, each stressing memory differently:
(i) \emph{information access and synthesis}, where memory holds what has already
been found and judged; (ii) \emph{personal assistance}, where memory holds a
persistent model of one user; (iii) \emph{procedural and environment-facing
tasks} such as coding, web navigation, and tool use, where memory holds
reusable, executable know-how; and (iv) \emph{long-horizon collaboration}, where
memory holds shared state across many sessions and possibly many participants.

Two observations follow. First, many of these are information tasks in the
conventional IR sense, but not all of them are: memory supports actions as well as
answers, extending beyond the scope of traditional IR. Second, the
application determines what counts as a good memory: a benchmark that measures
only fact recall in conversation will systematically favor systems designed for
one of these families and neglect the rest.

\subsection{Optimization: Learning What to Remember}
\label{sec:optimization}

The final component is the loop that uses feedback to improve these procedures. It
has two parts.

\emph{Reward modeling} asks what the feedback means. Feedback arrives as task
outcomes, user corrections, tool errors, and abandoned interactions, and must be
interpreted before it can be used. Determining its \emph{scope} is particularly difficult: when a
user rejects an itinerary, is the lesson specific to the current session, the
individual user, a cohort, or the entire population? This is the personalization spectrum of IR
--- session, individual, cohort, global --- reappearing as a credit-assignment
question. Assigning the wrong scope can cause a preference learned from one person to
leak into other users' agents. Extracting signal from noisy implicit
evidence is the problem that click models and unbiased learning to rank were
built for~\cite{chuklin2015click, joachims2017unbiased}.

\emph{Credit assignment} asks which memory decisions contributed to an outcome. A task
succeeds or fails after a long sequence of decisions about what to write,
retain, retrieve, and use. Improving the system requires attributing the outcome to points in
that chain and then updating the content of an item, the structure of the store,
or the policy that determines when to retrieve. Some of these updates can be
parametric, but most are not: they modify text and organization. This is a
central difficulty, and we return to it in Section~\ref{sec:optbeyond}.

\subsection{Existing Systems Are Special Cases}
\label{sec:special}

The framework should account for existing systems.
Table~\ref{tab:special} situates several representative systems within it. Our classification suggests
that these systems emphasize constants and utilization; support for reusable
functions, scoped operators, principled forgetting, and learned optimization is
less consistent. The underrepresented cells motivate parts of the research
agenda in Section~\ref{sec:agenda}.

\begin{table}[t]
  \caption{Representative agent-memory systems as special cases of the
  framework. $\bullet$ = central, $\circ$ = partial, blank = absent.
  Optimization: {\footnotesize N}one, {\footnotesize H}euristic,
  {\footnotesize L}earned.}
  \label{tab:special}
  \small
  \setlength{\tabcolsep}{4pt}
  \begin{tabular}{@{}lccccccc@{}}
    \toprule
    & \multicolumn{3}{c}{\textbf{Objects}} & \multicolumn{3}{c}{\textbf{Procedures}} & \textbf{Opt.} \\
    \cmidrule(lr){2-4}\cmidrule(lr){5-7}\cmidrule(lr){8-8}
    \textbf{System} & $C$ & $F$ & $H$ & Lrn & Fgt & Utl & \\
    \midrule
    RAG~\cite{lewis2020rag}                 & $\bullet$ &           &          &           &           & $\bullet$ & N \\
    MemGPT~\cite{packer2023memgpt}          & $\bullet$ &           & $\circ$  & $\bullet$ & $\circ$   & $\bullet$ & H \\
    Generative Agents~\cite{park2023generative} & $\bullet$ & $\circ$ & $\circ$ & $\bullet$ & $\circ$  & $\bullet$ & H \\
    Reflexion~\cite{shinn2023reflexion}     & $\circ$   & $\bullet$ &          & $\bullet$ &           & $\bullet$ & H \\
    Voyager~\cite{wang2023voyager}          & $\circ$   & $\bullet$ &          & $\bullet$ &           & $\bullet$ & H \\
    Mem0~\cite{chhikara2025mem0}            & $\bullet$ &           & $\circ$  & $\bullet$ & $\circ$   & $\bullet$ & H \\
    A-MEM~\cite{xu2025amem}                 & $\bullet$ & $\circ$   &          & $\bullet$ & $\circ$   & $\bullet$ & H \\
    PlugMem~\cite{yang2026plugmem}          & $\bullet$ & $\circ$   &          & $\bullet$ & $\circ$   & $\bullet$ & L \\
    \bottomrule
  \end{tabular}
\end{table}

%% file: sections/05_amm_vs_ir.tex
\section{Comparing AMM with Established IR}
\label{sec:vs}

Viewing AMM as a new form of IR problem invites an obvious question: if it is IR, what is new, and if it is new, why is it IR?
The four dimensions of the problem differ substantially in the degree to which they depart from established IR (Table~\ref{tab:vs}).

\begin{table}[t]
  \caption{Agentic memory management compared with established IR. The objects and
  the ultimate goal remain similar; the immediate consumer and the evaluation
  pipeline change more substantially.}
  \label{tab:vs}
  \small
  \setlength{\tabcolsep}{3.5pt}
  \begin{tabularx}{\columnwidth}{@{}>{\raggedright\arraybackslash}p{1.35cm}>{\raggedright\arraybackslash}X>{\raggedright\arraybackslash}X>{\raggedright\arraybackslash}p{1.2cm}@{}}
    \toprule
    & \textbf{AMM} & \textbf{Established IR} & \textbf{Change} \\
    \midrule
    Objects & all kinds of information, mostly interaction-derived and
    user-specific & all kinds of information, mostly authored and
    user-independent & degree \\
    \addlinespace[2pt]
    Users & agents, acting for people & people & kind \\
    \addlinespace[2pt]
    Goals & any agent activity performed on a user's behalf & mostly information
    access & scope \\
    \addlinespace[2pt]
    Evaluation & downstream task outcomes, more observable and replayable & proxy
    measures of ranking quality & pipeline \\
    \bottomrule
  \end{tabularx}
\end{table}

\subsection{Objects: A Difference of Degree}

Both fields manage many forms of information, and both must handle scale,
redundancy, staleness, and noise. The main differences are that memory content
is often \emph{produced through interaction} rather than authored independently,
and is frequently user-specific. This can make some established problems easier
(the source interaction may be logged; the collection may be small) and others
harder (there may be no crowd to aggregate over or anchor text, and personal
data entail obligations that a web crawl does not).

\subsection{Users: The Agent as a New Kind of User}

Here the change is substantial. IR's user models describe people: their needs,
their cognitive limits, their behavior under uncertainty. AMM's consumer is a
model. Without assuming humanlike cognition, we can still measure regularities
that play a role similar to behavioral biases in click modeling. A model attends
unevenly to long inputs, with strong position
effects~\cite{liu2024lost}. It may over-rely on retrieved content that
contradicts its parametric knowledge, or under-use content that is correct but
unconventionally formatted. Its propensity to retrieve at all depends on how
the task is framed.

This suggests an explicit research target: a \emph{utilization model}
for agents, the analogue of a click model, that predicts whether a given memory
item, in a given position and format, will actually change behavior. Such models
could serve the same three roles as click models: diagnostic
instruments, components of offline evaluation, and sufficiently differentiable
proxies to optimize against.

Treating the agent as a user also creates useful experimental opportunities. A model can be run
repeatedly, instrumented in detail, and replayed under controlled variations.
Experiments that are difficult to conduct with human participants can become practical. This
experimental advantage comes with the risk of non-stationarity: an agent's
behavior changes with every model version, so studies must identify the backbone
and, ideally, replicate their findings
across backbones. Recent analyses report that memory-system performance varies
substantially with the backbone~\cite{jiang2026anatomy}, so this is a
practical rather than a hypothetical concern.

\subsection{Evaluation: The Observability Inversion}
\label{sec:inversion}

The main consequence is methodological. IR's ultimate goal is to help someone
finish a task, but conventional IR systems often see only part of it. A search
engine observes a query, a ranking, and some clicks; it does not observe whether
the report was written, the trip was booked, or the bug was fixed. The field's
response became one of its central methodological conventions: hold the
downstream task constant by assumption, and measure the intermediate quantity we believe
predicts it --- relevance of a ranked list, evaluated on a reusable test
collection~\cite{sanderson2010test}. Proxy measures are especially important
when the final objective lies outside the system's view.

In AMM, this constraint weakens. The consumer of the
information is a system that we can instrument. We can observe which items were
retrieved, whether they entered the context, whether the agent's action changed
as a result, and whether the task succeeded. We can re-run the same model and
task with a different memory state, enabling a more controlled counterfactual
comparison than is usually possible after a person has experienced one
condition~\cite{balog2024usersim}. Evaluation can therefore connect component
behavior to downstream outcomes more directly. We call this shift the
\emph{observability inversion}: downstream use becomes more measurable, while
topical relevance becomes less informative on its own.

This argument requires two qualifications. First, end-to-end task success
\emph{confounds} memory quality with the agent's competence: a strong backbone
can compensate for a bad memory, and a weak one can waste a good one. Second,
outcome-only evaluation provides no diagnosis, so a failure cannot readily be
attributed to writing, retention, retrieval, or use.

The resolution is not to abandon component metrics but to change their status.
In established IR, a ranking metric is a \emph{proxy} for an objective that is
usually only partly observable. In AMM, more of the objective is observable, so
component measurements can become \emph{diagnostics} that explain an outcome. This
suggests evaluating memory systems as a stack:
(i) \emph{state quality} --- is the appropriate information stored at the right level of
abstraction, without contradiction;
(ii) \emph{utilization quality} --- is the stored item retrieved when it is
needed and does it change behavior when retrieved;
(iii) \emph{outcome} --- does the task succeed under a stated backbone; and
(iv) \emph{cost} --- what are the token, latency, and storage costs? Cost must
be reported alongside quality, since an accuracy improvement obtained at five
times the cost may not provide a net benefit. Existing benchmarks concentrate
on (iii) for a narrow slice of (i); few isolate (ii), and the validity of the
automated metrics used for (iii) is itself contested~\cite{jiang2026anatomy}.

%% file: sections/06_why_ir.tex
\section{Why IR Should Contribute}
\label{sec:whyir}

AMM is currently studied mostly in the NLP, ML, and agent communities,
and much of that work already draws on retrieval. Our claim is that IR offers
additional concepts and methods that should inform this work
more systematically.

\textbf{The goal is the same.} IR aims to help people satisfy information needs and complete tasks (Section~\ref{sec:retro}).
AMM aims to provide an agent with the information it needs to complete a task on a user's behalf.
AMM adds an intermediary while preserving the same objective.

\textbf{The techniques are shared.} Relevant methods include indexing,
retrieval and ranking, recommendation, clustering, query reformulation,
relevance feedback, compression, caching, and evaluation design. Structurally,
a memory system can be viewed as a small, personal, write-heavy, feedback-rich
retrieval system.

\textbf{The methodological experience is shared.} IR has shown how a proxy that
diverges from the objective can distort a system and how implicit feedback can
mislead when position and selection biases are ignored~\cite{joachims2017unbiased}.
It has also studied whether reported improvements survive comparison with fair
baselines and how shared benchmarks can either focus a field or narrow it
prematurely. These lessons can
inform the design of agent-memory systems and evaluations.

Two comparative advantages are worth naming. The first is \emph{methodological
rigor in studying users}. IR has spent decades building reproducible ways to study users and to
evaluate systems that serve them, including simulation as a first-class
methodology~\cite{balog2024usersim}. When the user becomes an agent, parts of
that methodological tradition transfer naturally and complement current memory
evaluation. The second is \emph{information management at
scale}, from index structures and update policies to modeling feedback from
noisy interactions.

We stress that this is an argument for contribution, not for ownership.
The shift toward generative information access is instructive: a vision articulated in IR~\cite{metzler2021rethinking} was realized mainly through LLMs developed outside the IR community (Section~\ref{sec:intro}).
AMM offers another opportunity for IR to shape a transition that draws directly on its core expertise.

%% file: sections/07_agenda.tex
\section{A Research Agenda}
\label{sec:agenda}

We follow the four directions introduced in Section~\ref{sec:intro}, and Section~\ref{sec:rosetta} additionally maps open AMM questions onto established IR problems.

\begin{table*}[t]
  \caption{A translation table between established IR problems and their agentic
  memory counterparts. The value of the mapping is not that established solutions apply
  unchanged, but that each row links an open question to a substantial body of
  prior research.}
  \label{tab:rosetta}
  \small
  \begin{tabularx}{\textwidth}{@{}>{\raggedright\arraybackslash}p{3.5cm}>{\raggedright\arraybackslash}p{5.1cm}>{\raggedright\arraybackslash}X@{}}
    \toprule
    \textbf{Established IR problem} & \textbf{Agentic memory counterpart} & \textbf{An open question} \\
    \midrule
    Indexing, index maintenance & memory organization and write policies &
    what structure supports retrieval by applicability and by scope, not only by content? \\
    Dynamic and temporal collections & updating and versioning constants &
    how should an agent decide that a remembered fact has expired rather than been contradicted? \\
    Relevance & usefulness for the next action &
    can usefulness be predicted before use, given that it depends on the agent, the task, and the format? \\
    Query understanding & constructing a memory query from agent state &
    what is the ``query'' when nobody typed one, and when should the agent issue one at all? \\
    Relevance feedback, click models & utilization feedback and credit assignment &
    what is the analogue of a click model for an agent that reads, and how do we debias it? \\
    Personalization granularity & scope of a learned preference &
    at which scope should a correction be applied, and how is over-generalization detected? \\
    Diversity and redundancy & deduplication, coverage, conflict resolution &
    when do two remembered items conflict rather than corroborate each other, and which should take precedence? \\
    Query performance prediction & predicting when memory helps or hurts &
    can an agent know in advance that retrieval will degrade its answer? \\
    Compression and efficiency & forgetting under a budget &
    what is the value of an item that has not yet been used? \\
    User simulation & agent behavior simulation for evaluation &
    how faithful must a simulated agent be for memory comparisons to hold? \\
    Privacy and fairness & deletion, leakage, and memory poisoning &
    what does a deletion guarantee mean once content has been lifted into a skill? \\
    \bottomrule
  \end{tabularx}
\end{table*}

\subsection{Toward Systematic Evaluation Protocols}
\label{sec:benchmarks}

IR made competing claims comparable through shared evaluation protocols, from the Cranfield experiments~\cite{cleverdon1967cranfield} to TREC~\cite{voorhees2005trec} and NTCIR~\cite{chen2025overview}, and progress in neural IR accelerated when large shared collections became available~\cite{nguyen2016msmarco}.
AMM benchmarks now exist~\cite{maharana2024locomo, wu2025longmemeval, ai2025memorybench}.
Most concentrate on conversational fact recall, although some have begun to include streams of user feedback~\cite{ai2025memorybench}; recent analysis also raises concerns about saturation, the validity of automated judges, backbone sensitivity, and incomplete cost reporting~\cite{jiang2026anatomy}.
Taking the Cranfield paradigm as a starting point, we suggest five desiderata for the next generation of AMM benchmarks.

\emph{Cover the object algebra.} A benchmark should test constants, functions,
and operators separately and in composition. Recalling a stated fact and
acquiring a workflow that transfers to a new site are different capabilities,
and a single aggregate score obscures which capabilities a system has acquired.

\emph{Test all three procedures.} Most benchmarks test utilization on a
preconstructed memory store. Learning (what does the system choose to store from a raw
episode?) and forgetting (what does it delete, and what breaks when it does?)
need first-class tasks with deliberately misleading, outdated,
or contradictory episodes, and deletion requests that must actually take effect.

\emph{Include feedback streams.} Optimization cannot be evaluated on static
data. A benchmark that hopes to measure whether a system \emph{improves} must
supply a stream of tasks with outcomes, corrections, and delayed rewards.

\emph{Make scope testable.} Multi-user, multi-session designs are needed to
detect a failure with serious consequences in deployment: a preference learned
from one user or one session leaking into contexts where it does not hold.

\emph{Report the backbone and the cost.} Results should be replicated across at
least two backbones, and token, latency, and storage costs reported alongside
quality, so that cost--quality trade-offs are visible rather than hidden.

\subsection{Reconnecting to Established IR}
\label{sec:rosetta}

New terminology can obscure useful connections to established work. For
example, contrastive objectives used in representation
learning~\cite{oord2018cpc} share structural features with ranking losses
developed earlier~\cite{cao2007listnet, burges2005ranknet}. AMM research on
deduplication, temporal validity, and feedback scope can likewise build on
related IR literature.

Table~\ref{tab:rosetta} is our attempt at a translation table. We do not claim
the established solutions transfer unchanged --- in most rows they do not, and the
interesting research lies in the mismatch. The mapping instead allows researchers
to build on a substantial body of prior work.

\subsection{Toward a Theory of Memory Utility}
\label{sec:theory}

As Section~\ref{sec:retro} notes, IR advanced through a combination of system building and foundational work.
Its foundational work provided explicit accounts of relevance and retrieval models derived from them~\cite{robertson1977prp, salton1975vector, ponte1998lm, robertson2009bm25}, giving later researchers clear assumptions to test and extend.
As data and model capacity grew, learned systems increasingly absorbed factors that these theories had enumerated explicitly~\cite{mitra2018neuralir}, which accelerated progress but provided less conceptual scaffolding.

Neither route currently serves AMM well.
An end-to-end learning approach is difficult to apply because the system is not one differentiable model but an assembly of prompts, stores, policies, and a foundation model whose retraining is out of reach for most researchers.
At the same time, system designs often rely on analogies to human memory --- humans have episodic memory, so the system has an episodic store --- without stating the conditions under which that design should help.
A productive next step is to state hypotheses about memory systems, derive observable consequences, and test them.

We therefore state three candidate principles as conjectures to be tested
rather than as established results.

\textbf{C1: Memory Utility Principle.} A memory system should prioritize items
for retention and retrieval according to their expected marginal contribution
to future task reward, net of the associated costs --- storage, retrieval,
context occupancy, and interference with other items. This formulation
deliberately mirrors the probability ranking
principle~\cite{robertson1977prp}, and it inherits both its appeal and its
weaknesses: the expectation is over an unknown future task distribution, and
marginal contribution is not additive across items.

\textbf{C2: Abstraction trade-off.} Lifting constants into functions and
functions into operators increases transfer but also increases the risk of
misapplication. For a given task distribution, there may therefore be an optimal
level of abstraction. One testable corollary is that systems that aggressively
summarize episodes into general rules should outperform verbatim storage on
task distributions with high repetition, and underperform it on heterogeneous
ones.

\textbf{C3: Minimal scope.} A correction should be applied at the narrowest
scope that explains it: the system should prefer a session-specific scope to a
user-specific one, and a user-specific scope to a cohort-wide one. The scope should widen only with
repeated, independent evidence.
Violating this principle offers one plausible explanation for personalization
failures in which an agent overgeneralizes a local correction.

Each conjecture is testable with the benchmark designs of
Section~\ref{sec:benchmarks}. Evidence that refutes any of them would still sharpen
the research agenda.

\subsection{Hybrid Parametric and Non-Parametric Optimization}
\label{sec:optbeyond}

A memory system can improve from experience through parametric updates to model weights and through non-parametric updates to the content and structure of its store.
For non-parametric updates, something must play the role that gradient descent plays for parametric models.
Today that role is usually filled by asking a language model to revise its own notes, an approach formalized as textual ``gradients''~\cite{yuksekgonul2024textgrad}.
It is often effective in practice, but current accounts provide limited guidance about when repeated revision improves task performance, whether the process converges, or how to distinguish a useful update from mere rephrasing.
These questions mark a substantial theoretical gap.

We see four concrete directions. First, \emph{characterize non-parametric
updates}: under what conditions does iterated textual revision converge, and can
guarantees of monotonic improvement be obtained under assumptions about feedback
quality? Second, \emph{credit assignment}: attributing a delayed, sparse task
outcome to decisions along a trajectory of writing, retention, retrieval, and use is a
counterfactual estimation problem, for which IR's work on counterfactual and
online learning to rank~\cite{joachims2017unbiased} provides a relevant toolkit.
Third, \emph{joint parametric and non-parametric optimization}: some knowledge
belongs in weights, whereas other knowledge belongs in text. The boundary is currently determined by
convenience, and a principled account of consolidation --- what to compile into
parameters, when, and how to reverse it --- is missing. Fourth,
\emph{optimizing structure, not only content}: the organization of the store is
itself a decision variable, and changing it changes what can be retrieved.

\subsection{Studying the Agent, and Keeping the Human in the Loop}
\label{sec:agentuser}

Finally, the user view of Section~\ref{sec:retro} should be applied to the new user.
Agent interaction logs invite analyses analogous to those applied to query logs~\cite{silverstein1999analysis}: what agents retrieve, when they retrieve it, how often retrieval proves unhelpful, and what a trajectory of memory operations looks like when a task goes wrong.
Such analyses could lead to a taxonomy of agent information needs, in the way that Broder's taxonomy characterized web search~\cite{broder2002taxonomy}.
Memory presentation deserves the attention that result-page design once received, since format, granularity, ordering, and attribution all change whether an item is used.
Agents can also be trained to search their own memory more effectively, which is itself an IR problem.

Human control remains essential. Because memory is personal, persistent, and
consequential, people need to see, correct, and delete what an agent has
remembered about them. Deletion must remain meaningful even after the content
has been lifted into a skill or consolidated into parameters. Memory is also
an attack surface: poisoning of memory stores~\cite{chen2024agentpoison} and
membership inference against them~\cite{chen2026membership} are emerging
concerns. These are governance
questions, but they are also retrieval questions, and they will not be answered
by systems work alone.

%% file: sections/09_conclusion.tex
\section{Conclusion}
\label{sec:conclusion}

Salton's definition of IR has outlived the technology for which it was written and remains applicable to this transition.
The activities it names --- structuring, analyzing, organizing, storing, searching, and retrieving --- remain the same, but the consumer for which they are performed has changed, and AMM adds one more: the \emph{use} of information by the agent itself.
When an agent acts on a person's behalf, the information that matters is increasingly the information the agent keeps for itself, and managing it well is what allows an assistant to improve with use rather than start afresh in every session.

AMM can therefore be viewed as a new form of IR problem, although it is currently approached mainly through system building and analogies to human memory.
IR can help build stronger foundations for it by developing systematic evaluation protocols, stating testable principles of memory utility, designing optimization methods that combine parametric and non-parametric updates, and keeping humans in the loop.
The field built such foundations for library collections and later for the web, starting from the vision of the memex as an external memory for a person~\cite{bush1945memex}.
The next memory to organize is the one inside the agent.